\input psfig.tex
\documentstyle[12pt,pageset]{article}
\unipage
\def\reference{\parskip 0pt\par\noindent\hangindent 0.5 truecm}

\begin{document}
%
%
\title{On photohadronic processes in astrophysical environments}
%


\author{A.~M\"ucke$^1$ \and
 J.P.~Rachen$^2$ \and
 Ralph~Engel$^3$ \and
 R.J.~Protheroe$^1$ \and
 Todor~Stanev$^3$
} 

\date{}
\maketitle

{\center
$^1$ University of Adelaide, Dept. of Physics \& Math.Physics, Adelaide, SA 5005, Australia\\ amuecke@physics.adelaide.edu.au, rprother@physics.adelaide.edu.au\\[3mm]
$^2$ Pennsylvania State University, Dept. of Astronomy, University Park, PA 16802, USA\\jorg@astro.psu.edu\\[3mm]
$^3$ Bartol Research Institute, University of Delaware, Newark, DE 19716, USA\\eng@lepton.bartol.udel.edu, stanev@bartol.udel.edu\\[3mm]
}

%
\begin{abstract}
  We discuss the first applications of our newly developed Monte Carlo event
  generator SOPHIA to multiparticle photoproduction of relativistic protons
  with thermal and power law radiation fields.  The measured total cross
  section is reproduced in terms of excitation and decay of baryon resonances,
  direct pion production, diffractive scattering, and non-diffractive
  multiparticle production. Non--diffractive multiparticle production is
  described using a string fragmentation model.  We demonstrate that the
  widely used `$\Delta$--approximation' for the photoproduction cross section
  is reasonable only for a restricted set of astrophysical applications.  The
  relevance of this result for cosmic ray propagation through the microwave
  background and hadronic models of active galactic nuclei and gamma-ray
  bursts is briefly discussed.

\end{abstract}

{\bf Keywords:}
nuclear reactions -- elementary particles -- radiation mechanisms: non-thermal

\bigskip

%
%
\section{Introduction}
The energy spectrum of cosmic rays extends to energies above $10^{20}$\,eV.
Such particles are most likely extragalactic and propagate through many tens
of Mpc before reaching the Earth. The main energy loss mechanism for nucleons
in this energy range is photomeson production on the cosmic microwave
background radiation (CMBR). This process 
has an energy threshold of $1.08$ GeV in
the center of momentum frame (CMF) of the interacting particles.
It causes the distortion of the proton spectrum above $3\times 10^{19}$
eV during propagation, known as the Greisen--Zatsepin--Kuzmin cutoff 
(Greisen 1966, Zatsepin \&
Kuzmin 1966; see Protheroe \& Johnson 1996 for additional references).
In environments harbouring dense radiation fields with higher photon
energies the particle production threshold is reached at lower proton
energies, e.g.
${\sim}\,10^{16}$\,eV for ambient photon energies of 10 eV. If protons
are accelerated in such energetic astrophysical objects, photopion production
and subsequent pion decay would lead to the emission of energetic
$\gamma$-rays and neutrinos. This may be observable from jets in active
galactic nuclei (AGN) (Mannheim
1993; Protheroe 1996) and gamma-ray bursts (GRB) 
(Waxmann \& Bahcall 1997; Vietri
1998a,b; B\"ottcher \& Dermer 1998; Rachen \& Meszaros 1998).

The prominent $\Delta(1232)$--resonance near the threshold
has often been used to construct approximate pion production cross sections
(referred to as the `$\Delta$--approximation' hereafter)
and to determine gross features like $\gamma$-ray-to-neutrino energy yields,
proton inelasticities, etc. which enter the relevant astrophysical
calculations.  In this paper we discuss first results using a new Monte-Carlo
event generator for photohadronic interactions of relativistic protons in
radiation fields of astrophysical origin, which includes all relevant
interaction processes and is based on models
and data available from particle physics.  

Our event generator, SOPHIA (\underline{S}imulation \underline{O}f
\underline{P}hoto-\underline{H}adronic \underline{I}nteractions
in \underline{A}stro\-physics), has been extensively tested on all fixed target
and collider experiment data available to us (M\"ucke et al.\ 1998a). A
comparison of the results from SOPHIA with the 
$\Delta$--approximation emphasizes the importance of including all
interaction processes
in calculations of expected astrophysical
signals. Detailed estimates of such signals will be subject of future
work. This paper concentrates on photopion production features that may
be important for astrophysical processes which involve hadronic reactions.
In Sec.~2 we summarize the physics implemented in SOPHIA and
contrast it with the $\Delta$--resonance approximation.  
Various
astrophysical applications of SOPHIA are discussed in Sec.~3 
by considering power-law
and thermal photon seed spectra. A summary is given in Sec.~4.

 
\section{Cross section and kinematics}

Consider a relativistic proton with energy $E_p=\gamma m_p$ and rest mass
$m_p$, which interacts with a photon of energy $\epsilon$ at an angle
$\theta$. The square of the total CM frame energy of the interaction is given
by
\begin{equation}
s = m_p^2 + 2E_p\epsilon(1-\beta\cos{\theta}) = m_p^2 +
2m_p\epsilon^\prime,
\end{equation}
where
$\beta=\sqrt{1-\gamma^{-2}}$ is the velocity of the proton in terms of the
velocity of light and $\epsilon^{\prime}$ the photon energy in the nucleon
rest frame (NRF). In SOPHIA, proton and neutron
induced interactions are distinguished -- for simplicity we quote only
proton interactions in the following.

The partial cross sections for
resonance excitation, direct (non-resonant) single-pion production,
and diffractive scattering are
determined by fits to exclusive data (Fig.~1).  We consider the 9 most
important resonances ($\Delta^+(1232)$, $N^+(1440)$, $N^+(1520)$,
$N^+(1535)$, $N^+(1650)$, $N^+(1680)$, 
$\Delta^+(1700)$, $\Delta^+(1905)$ and $\Delta^+(1950)$) 
and use the Breit-Wigner formula
together with their known properties of mass, width and decay branching
ratios to determine their contribution to the individual interaction
channels. The direct channel is defined as the residual, non-resonant
contribution to the channels $p\gamma \to n\pi^+$, $p\gamma \to
\Delta^{++}\pi^-$, and $p\gamma \to\Delta^{0}\pi^+$. The method is
described in detail in Rachen (1996).
Note that, although the
$\Delta(1232)$-resonance has the highest cross section at low interaction
energies, the direct channel dominates near threshold.
The direct channel can be important for proton interactions in soft
photon spectra since it produces exclusively charged pions.  
Diffractive scattering is due to the coupling of the photon to the 
vector mesons $\rho^0$ and $\omega$, which are produced at very high
interaction energies with the ratio
9:1 and a cross section proportional to the total cross section. Finally, the
residuals to the total cross section are fitted by a simple model and treated
as statistical multipion production which is simulated by a QCD string
fragmentation model (Andersson et al. 1983).  
After the decay of all intermediate states, considering
basic kinematical relations and accelerator data on rapidity distributions,
the resulting distributions of protons, neutrons and pions in the source
determine the measurable astrophysical quantities, for example, the cosmic ray,
neutrino and $\gamma$-ray emission. Details of the simulation techniques
are described in M\"ucke et al.\ (1998a).
\begin{figure}[htb]
\centerline{\psfig{figure=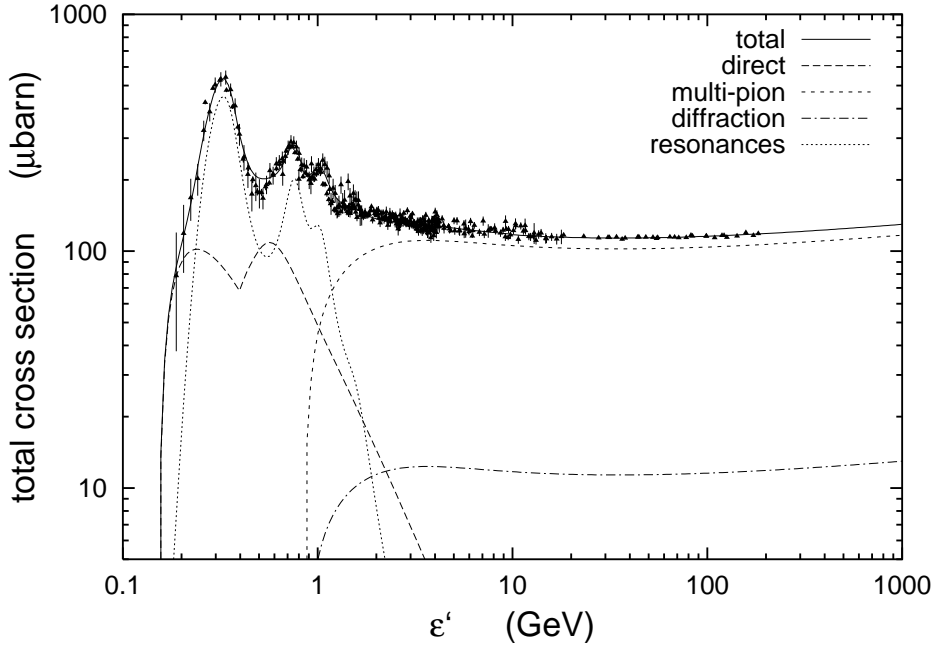,height=9cm}}
\caption[fig1]{The total $p\gamma$ cross section with the 
contributions of the baryon resonances considered in this work,
the direct single-pion 
production, diffractive scattering, and the multipion production 
as a function of the photon's NRF energy (1 $\mu$barn = $10^{-30}$
cm$^2$). Data are from Baldini et al.\ (1988). 
}
\end{figure}

In the $\Delta$--approximation (Stecker 1973, Gaisser et al.\ 1995) the
total cross section is given as $\sigma_{\Delta} = 500 ~\mu\rm{barn}
\, \Theta(\sqrt{s} - m_\Delta+\Gamma_\Delta/2) \cdot \Theta(m_\Delta +
\Gamma_\Delta/2-\sqrt{s})$, where $m_\Delta$ = 1.232~GeV is the mass
and $\Gamma_\Delta = 0.115$ GeV is the width of the
$\Delta(1232)$--resonance,
and $\Theta$ is the Heaviside step function.  The
$\Delta$--approximation uses the branching ratios of the
$\Delta^+(1232)$-resonance to determine the number ratio
$\pi^0$ to $\pi^+$ of 2:1. Photohadronic $\nu$-production is the result of the
decay of charged secondary pions ($\pi^+\rightarrow
e^+\nu_{\mu}\bar\nu_{\mu}\nu_e$, $\pi^-\rightarrow
e^-\nu_{\mu}\bar\nu_{\mu}\bar\nu_e$).  Gamma rays are produced via neutral
pion decay ($\pi^0\rightarrow\gamma\gamma$) and synchrotron/Compton emission
{}from the resulting relativistic leptons.  
In the $\Delta$--approximation
this leads to a ratio of the energy content in
$\gamma$--rays to neutrinos $\sum E_\gamma:\sum E_\nu \equiv {\cal
E_\gamma}:{\cal E_\nu} = 3:1$.
The decay kinematics of this resonance decay predicts a 
nucleon inelasticity $K_p = \Delta
E_p/E_p\approx 0.2$.

\section{Astrophysical applications}

 In this section we illustrate the importance of using the complete
 photohadronic cross section and final state composition and kinematics
 for a number of astrophysical applications. In particular,
 we compare the average proton inelasticity in a photopion production
$K_p$
 and the ratio
 ${\cal E}_\gamma/{\cal E}_{\nu}$ 
 from simulations with SOPHIA to the $\Delta$--approximation. 
We assume that due to radiative processes electrons convert all their
kinetic energy into photons, and so $e^{\pm}$ are counted in ${\cal E_\gamma}$.
The total neutrino energy
 ${\cal E}_{\nu}$ is the sum of the energies of $\nu_e$, $\nu_\mu$ and their
 antiparticles.

\subsection{Power--law seed photon spectra}

 In order to illustrate the contribution of the different interaction
processes in 
various astrophysical
 environments we convolute the cross section with proton
and seed photon spectra.
 Proton spectra as well as non-thermal seed photon spectra can
 be well approximated by power laws. Figure~2 shows the distribution of
 the number of interactions with CMF energy squared, $s$, for 
power--law seed photon
 spectra $n_\epsilon \sim \epsilon^{-\alpha_\gamma}$
 and a proton spectrum $n_E \sim E_p^{-\alpha_p}$ 
 with index $\alpha_p = 2$. Spectra with $\alpha_p = 2$
 are typical for shock accelerated particles. Shock acceleration is
believed to be the dominant process for acceleration of galactic cosmic
rays (Jones \& Ellison 1991). It may also be present in extragalactic 
 particle accelerator 
 (Biermann \& Strittmatter 1987),  in particular AGN
 (e.g. Stecker et al.\ 1991, Mannheim 1993, Protheroe 1996) 
 and in GRBs (e.g. Waxman 1995, Vietri 1998). 

\begin{figure}[htb]
\centerline{\psfig{figure=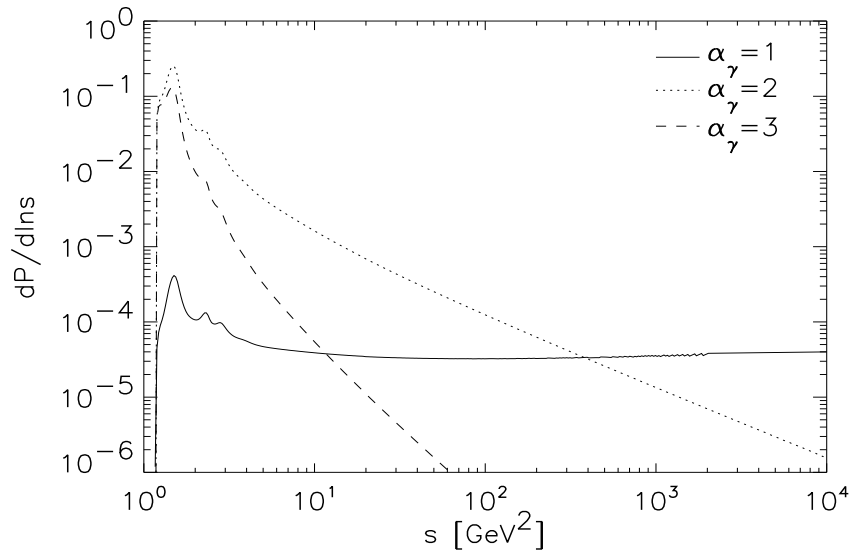,height=8cm,width=12cm}}
\caption[fig2]{
Interaction probability distribution as function of the squared CMF
energy
with index $\alpha_{\gamma}$ and proton spectrum $n_E \sim E^{-2}$. 
}
\end{figure}

It has been proposed that the high energy gamma ray emission
 from blazar jets as observed in the GeV to TeV energy range results
 from photopion production in a proton-electron plasma where the relativistic
 electrons provide a synchrotron seed photon field for the high energy
 protons (``Proton Blazar'' model, Mannheim 1993).
 The synchrotron photon spectrum is described by a power law 
 which is often observed in jets of AGN. 
 Photopion production is followed by electromagnetic cascades reprocessing
 the injected power of the pion decay products (Mannheim et al.\ 1991). 
Therefore, Fig.~2 may approximate
 the situation in AGN jets. While for steep photon spectra the prominent
 resonance region near threshold is the dominating contributor to the
 interaction rate, the high energy region of the cross section gains
 importance for flat seed photon spectra.
 Radio spectra with
$\alpha_\gamma\approx1$ ($F_\nu \propto {\rm const.}$) up to a break 
frequency $\nu_b$ are,
 for example, observed in flat spectrum radio quasars (FSRQs) and BL~Lac
 objects. Such spectra are believed to originate from a superposition of
 several self-absorbed synchrotron components (Cotton et al.\ 1980,
 Shaffer \& Marscher 1979). Above $\nu_b$ the spectrum is
 loss dominated (i.e., the cooling time scale is shorter than the
 dynamical time scale) and steepens to $\alpha_{\gamma} \approx 2$. 
 Recent observations of blazars have revealed synchrotron peaks ranging from the
 optical/IR-band up to UV/soft X-rays, especially in the
 flaring state (see, e.g., Fossati et al.\ 1998). 
 Thus flat power law photon spectra exist in AGN jets and would shift
the average CMF energy of photopion production towards high values.
 For protons  with energies below
 $10^{16}$\,eV only the infrared and higher frequency photons
 are relevant for photopion production due to the threshold condition.
 For blazars with a synchrotron peak at rather low frequencies the
target
 photon spectrum for photomeson production may be rather steep,
 and photohadronic interactions would mainly occur near the threshold.
 

In both, flat and steep photon spectra, the simulations with SOPHIA show that
 the $\gamma$-ray-to-$\nu$-energy ratio is  
${\cal E}_{\gamma}/{\cal E}_{\nu} \approx 1$.
The difference with the $\Delta$--approximation (${\cal
E}_{\gamma}/{\cal E}_{\nu} = 3$)
 is due to the contribution of the
 second resonance and the multipion production regions in the case of
flat photon spectra. In steep spectra, the difference is
explained by the influence of the immediate threshold region, where the
direct channel dominates over the Delta resonance (M\"ucke et al.
1998b). 
Therefore, the $\gamma$-ray and
neutrino outputs are approximately equal.
 This may have significant impact on estimates
 of the predicted neutrino fluxes from AGN jets which are usually normalized
 to the observed $\gamma$-ray luminosity. 

The absolute $\gamma$-ray and neutrino energy outputs depend on the
total proton energy loss $K_p E_p$. For photon power law indices
$\alpha_\gamma
\geq 2$ we find $K_p \approx 0.2$, as in the $\Delta$-approximation.
  The inelasticity increases rapidly for flatter photon spectra to reach
$K_p \approx 0.6$ for $\alpha_\gamma = 1$.
 This shows that the $\Delta$-approximation
 underestimates significantly the fractional energy loss of the incident
 nucleon in environments with flat seed photon spectra, where protons
 cool much faster.

Proton energy losses, however, limit the maximum possible proton
 energy which can be achieved in AGN jets (Biermann \& Strittmatter 1987).
 Setting the time scales for proton acceleration and energy loss equal,
Mannheim (1993) obtained
 cutoff energies of $\sim 10^{16}$\,eV. 
This estimate is based on a mean cross section
 for photopion production $\langle K_p \sigma_{p\gamma}\rangle=50\mu\rm{barn}$
 and a mean inelasticity $\langle K_p\rangle=0.25$. Noting that for flat seed
photon spectra SOPHIA simulations give proton inelasticities
about a factor of 2 larger than Mannheim (1993), we expect slightly 
lower maximum proton energies.
Similar corrections could apply to any proton acceleration model
where the maximum proton energy is limited by photohadronic
interactions.
In contrast, models 
in which Larmor radius constraints and finite acceleration time scales limit
the proton energies, would not be affected (e.g. proton acceleration in
hot spots of radio galaxies, Rachen \& Biermann 1993).

\begin{figure}[htb]
\centerline{\psfig{figure=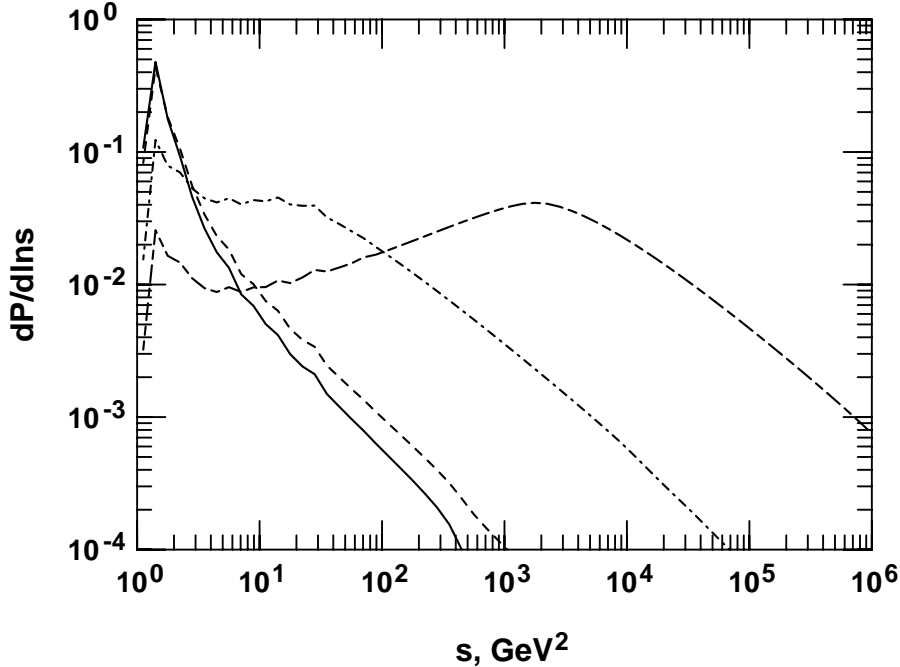,height=9cm,width=12cm}}
\caption[fig3]{
Interaction probability distribution as function of the squared CMF
energy
 for a broken power-law seed photon spectra as observed in GRBs ($n_{\epsilon}
 \sim \epsilon^{-2/3}$ for $10^{-8}eV\leq\epsilon\leq 1$ keV and $n_{\epsilon}
 \sim \epsilon^{-2}$ for 1 keV$\leq\epsilon\leq$100 keV and proton spectrum
 $n_E \sim E^{-2}$ with $E_{\rm{min}}\leq E \leq 10^{21}$\,eV and different minimum
 energies $E_{\rm{min}}=10^{12}$\,eV (solid line), $10^{14}$\,eV (dashed line), $10^{16}
 $\,eV (dashed-dotted line) and $10^{18}$\,eV (dashed-dashed line).
}
\end{figure}

An ideal environment for proton acceleration may be found in
 gamma-ray bursts which have a broken power-law photon spectrum
steepening from $\alpha_\gamma \approx 1$ to $\alpha_\gamma \approx
2$ at keV energies (in the comoving frame). 
Some GRB models involve the acceleration of
 ultrahigh energy cosmic rays.
 Cascades initiated by photopion
 production follow (Waxman \& Bahcall 1997, Vietri 1998, B\"ottcher
 \& Dermer 1998, Rachen \& Meszaros 1998). Figure~3 shows the probability
of photoproduction interaction at  squared CMF energy $s$
for a $E^{-2}$-proton spectrum with
 a photon spectrum with $\alpha_\gamma = {2/3}$ below 1 keV and
 $\alpha_\gamma =2$ above that energy. The curves are calculated
 for different parts of the proton spectrum, extending from minimum
energies of $10^{12}{-}10^{18}$\,eV to a high energy
 cutoff of $10^{21}$\,eV.
While for a proton spectrum with
 $E_{\rm min}=10^{12}$\,eV most interactions occur 
on the $\epsilon^{-2}$-part of the
 photon field, for higher $E_{\rm{min}}$ proton-photon interactions 
involve the flat
part of the seed photon spectrum leading to 
 higher average $s$ values.
Consequently, a $\gamma$-ray-to neutrino energy ratio of $1$ 
and
 inelasticities of the order of $0.5{-}0.7$
 are expected for the production of the highest neutrino and secondary
 photon energies (M\"ucke et al., 1998b). 

Recently, Vietri (1998a,b) has claimed  that in situ
photoproduced neutrinos in the blast-waves  associated with GRBs
 can reach energies up to ${\approx}10^{19}$\,eV. In order to calculate
the maximum neutrino energy, he assumed that individual neutrinos
typically carry 5\% of the incident proton energy which is derived from
the $\Delta$-approximation.
 However, SOPHIA
 simulations show that this value is rather about 1\% due to the
increasing neutrino multiplicity, constraining the
maximum neutrino energies to lower values (M\"ucke et al.\ 1998b).

Since the average interaction energies are well beyond
the resonance region in GRB we expect that detailed calculations with
SOPHIA will change significantly the predictions of hadronic GRB models.

\subsection{Thermal seed photon spectra}

By contrast to power-law photon spectra, thermal (black body) photon
fields are strongly concentrated about a characteristic energy $k T$.
This concentration of the seed photons leads to an emphasis of one
specific photoproduction energy range.
 We consider the interactions of protons with different power law
spectra with photons
 from $T = 2.73$ K and
 $k T = 10$\,eV blackbody spectra.
 The former case is of particular astrophysical interest for cosmic ray
 (CR) propagation through the cosmic microwave background.
 The latter
 case may approximately describe the situation in AGN jets with the
 accretion disk photons being the target for proton photon interactions.
We show the interaction probability as a function of  $s$ in Fig.~4. 
In the case of CR transport through the CMBR 
  the $\Delta(1232)$-resonance region  plays the dominant role. 
 SOPHIA simulations give a photon-to-neutrino energy ratio
 ${\cal E}_{\gamma}/{\cal E}_{\nu} \approx 1.7$ and inelasticities 
$\approx 0.2$.
 We note that the neutrino component
 dominates directly at the threshold with 
${\cal E}_{\gamma}/{\cal E}_{\nu} \approx 1/2$ due
 to the dominance of the direct channel.

 Recent AGASA measurements seem to indicate an extension of the CR spectrum
 beyond $10^{20}$\,eV (Takeda et al.\ 1998). 
It is generally assumed that acceleration scenarios are limited to
maximum energies of $10^{21}$\,eV. If the CR proton spectrum extends 
 above $10^{21}$\,eV, one may have to consider so called top--down (TD)
 models where the highest energy cosmic rays are decay products of
 GUT scale ultraheavy ($10^{25}$\,eV) particles 
(e.g., Hill 1983; Sigl et al.\ 1994; Protheroe \& Stanev 1996).
While for TD models the
 photon-to-neutrino energy ratio
 would reach unity and proton inelasticities can rise up to $0.6$,
in case of acceleration scenarios the $\Delta$-approximation describes
sufficiently well the proton propagation through the CMBR.

In previous Monte Carlo calculations of cosmic ray pion photoproduction
interactions (e.g. Protheroe \& Johnson 1996), an earlier photoproduction
event generator based on the observed inclusive cross sections for $p\gamma
\to \pi^0p$, and for $p\gamma \to \pi^+n$, together with a semi-empirical
model for multi-pion production (with equal numbers of $\pi^+$, $\pi^-$ and
$\pi^0$ produced in the central region, and charge ratios in the fragmentation
regions based on naive quark model arguments; Szabo \& Protheroe 1994) was
used.  The results from this previous event generator are not too different,
and give ${\cal E}_{\gamma}/{\cal E}_{\nu} \approx 1.6$ and inelasticity
${\approx}0.2$ at the peak of the resonance, and ${\cal E}_{\gamma}/{\cal
  E}_{\nu} \approx 1.0$ and inelasticity ${\approx}0.55$ in the extreme
multipion region.

For photopion production in an isotropic UV radiation field, the high energy
and the resonance regions of the cross section are of equal importance (see
Fig. 4). Thermal radiation peaking in the UV-range of the electromagnetic
spectrum is observed from some radio-loud AGN (for example 3C\,273). Therefore
a hadronic blazar model has been proposed in which protons accelerated in some
regions of the jet interact with accretion disk photons (``Accretion Disk
Proton Blazar Model'', Protheroe 1997, 1998).  In this model the protons,
being charged, are isotropised in the jet frame and readily interact with the
almost black-body photons coming from the inner part of the accretion disk
near the base of the jet, provided the disk is sufficiently luminous or the
emission region is not too far away from the disk. The model is tuned such
that the resulting $\gamma$-rays, which are relativistically beamed along the
jet direction, undergo a pair-synchrotron cascade in the accretion disk
radiation and jet magnetic field such that the emerging radiation can
reproduce features of the observed $\gamma$-ray emission of blazars.  The
physics implemented in SOPHIA allows for a detailed investigation of the
parameter space of this model.
%
%
\begin{figure}[htb]
\centerline{\psfig{figure=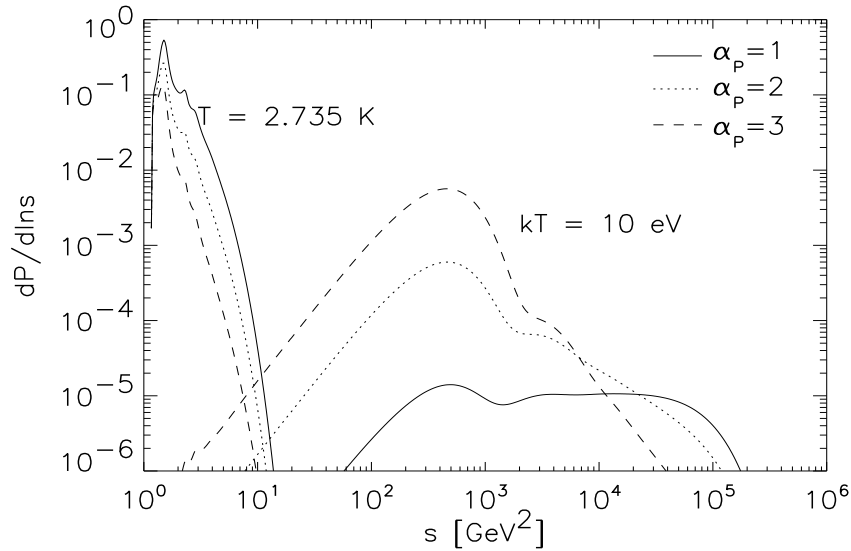,height=8cm,width=12cm}}
\caption[fig4]{
Interaction probability distribution as function of the squared CMF
energy for
black body seed photon spectra 
and proton spectra with index $\alpha_p$ between $10^{6\ldots12}$GeV.
}
\end{figure}

\section{Conclusions}

We use our recently developed Monte Carlo code for photohadronic processes
(SOPHIA) to compare predictions of astrophysically relevant quantities with
the expectations from the widely used `$\Delta$-approximation'.  We discuss
proton inelasticities and $\gamma$-ray-to-neutrino energy ratios in the
framework of astrophysical environments like AGN, GRBs and for the case of CR
propagation.  SOPHIA uses the full information of the cross section and the
final state particle composition and kinematics of the interaction processes
as provided by particle physics. We show that the $\Delta$-approximation is a
reasonable approximation only for a restricted set of applications.  Several
astrophysical cases are discussed:
\begin{itemize}
\item Photohadronic interactions of high energy nucleons in typical radiation
  fields of GRBs, which may be responsible for the highest energy
  $\gamma$-rays and neutrinos, occur mainly at high CMF energies.  Nucleon
  inelasticities of $\approx 0.5{-}0.6$ are expected in this case
  whereas the $\Delta$-approximation gives only ${\approx} 0.2$.  The
  $\gamma$-ray and neutrino energy outputs are approximately equal.
\item 
High-energy $\gamma$-ray emission from blazar jets may be
 explainable by photopion production of relativistic protons bathed in a
 synchrotron or thermal (UV-photons from the accretion disk, IR-photons
{}from the molecular torus) radiation field (``Proton Blazar'' models). For a
flat-spectrum
 seed photon field photohadronic interactions with high CMF energies
 become important. 
SOPHIA simulations show that protons may cool much faster than in the
$\Delta$-approximation because of the growth of the inelasticity $K_p$.
 The $\gamma$-ray-to-neutrino energy ratio may approach unity for both, flat and
steep seed photon spectra, whereas
 many models use ${\cal E}_{\gamma}/{\cal E}_{\nu}=3$. This would affect
estimates of the predicted neutrino flux. 
\item 
Interactions of ultra-high energy cosmic rays  
propagating through the microwave background mainly
 happen in the region of the $\Delta(1232)$-resonance for
energies
 ${<}10^{21}$\,eV. Above that energy the effects of the second
 resonant and multiparticle regions increase the fractional energy loss
and alter the final shape of the cosmic ray spectrum.
\end{itemize}
Detailed calculations addressing photomeson production in the
 above mentioned astrophysical environments will be reported in
forthcoming publications.





\section*{Acknowledgements}

 The work of AM and RJP is supported by the Australian Research Council.
 RE and TS acknowledge the support by the U.S. Department of Energy
under
 Grant Number DE FG02 01 ER 40626. The work of TS is also supported in
 part by NASA NAG5--7009. The work of JPR is supported by NASA
NAG5-2857.
 TS thanks the University of Adelaide for hospitality during his visit,
 which was funded in part by the International Visitor Program of the
 Special Research Center for Theoretical Astrophysics, University of
 Sydney, when this work was started.

\section*{References}






\reference Andersson, B., et al., 1983,  Phys. Rep. 97, 31
\reference Biermann, P.L. \& Strittmatter, P.A., 1987, ApJ, 322, 643 
\reference B\"ottcher, M. \& Dermer, C.D. 1998, astro-ph/9801027
\reference Cotton, W.D., Wittels, J.J., Shapiro, I.I. et al., 1980, ApJL, 238, L123
\reference Fossati, G., et al., 1998, MNRAS, in press.
\reference Gaisser, T.K., Halzen, F., Stanev, T. 1995, Phys.Rep., 258, no.3
\reference Greisen, K., 1966, Phys. Rev. Lett., 16, 748
\reference Hill, C.T., 1983, Nucl. Phys. B, 224, 469
\reference Jones, F.C., \& Ellison, D.C., 1991, Space Sci. Rev., 58, 259
\reference Mannheim, K., Kr\"ulls, W. M. and Biermann, P. L., 1991, A\&A,
251, 723
\reference Mannheim, K. 1993, A\&A, 269, 67
\reference M\"ucke, A., et al.\ 1998a, Comp. Phys. Comm, in preparation
\reference M\"ucke, A., et al.\ 1998b, in preparation
\reference Protheroe R.J. \& Johnson, P.J. 1996, Astroparticle Phys., 4, 253
 and erratum 5, 215
\reference Protheroe R.J. \& Stanev T., 1996, Phys. Rev. Lett., 77, 3708
\reference Protheroe, R.J., 1997,
        in Accretion Phenomena and Related Outflows, IAU
        Colloq. 163, Port Douglas, July 1996,
        ed. D.T. Wickramasinghe et al., ASP Conf. series,
        Vol. 121, 585
\reference Protheroe, R.J., 1998,
        Publications of the Astron. Soc. Aust., to be submitted
\reference Rachen, J.P., 1996, PhD-thesis, MPIfR Bonn, Germany\\
(http://www.astro.psu.edu/users/jorg/PhD)
\reference Rachen, J.P. \& Biermann, P.L., 1993, A\&A, 272, 161
\reference Rachen J.P., \& Meszaros, P. 1998, astro-ph/9802280, Phys.
Rev. D in press.
\reference Baldini, A., Flaminio, V., Moorhead, W. G., Morrison, D.R.O.,
1988, in: Landold-B\"ornstein, New Series, Vol. I/12b, 
edited by H. Schopper, Springer-Verlag and references therein.
\reference Shaffer D.B.\& Marscher, A.P., 1979, ApJL, 233, L105
\reference Sigl, G., Schramm, D. N. \& Bhattacharjee, P., 1994.
Astropart.Phys., 2, 401.
\reference Stecker, F.W., 1973, Astrophysics and Space Science, 20, 47
\reference Stecker, F.W. et al., 1991, Phys. Rev. Lett., 66, 2697 and erratum 69, 2738 (1992)
\reference Szabo, A.P., and R.J. Protheroe, 1994, Astroparticle Phys., 2, 375 
\reference Takeda, M. et al, 1998, Phys. Rev. Lett., 81, 1163.
\reference Vietri, M. 1995, ApJ, 453, 883. 
\reference Vietri, M. 1998a, ApJ, in press.
\reference Vietri, M. 1998b, Phys. Rev. Lett., 80, 3690. 
\reference Waxman, E. J., 1995, Phys. Rev. Lett., 75, 386
\reference Waxman, E. \& Bahcall, J., 1997, Phys. Rev. Lett., 78, 2292
\reference Zatsepin, G.T. \& Kuzmin, V.A., 1966, JETPh Lett, 4, 78

\end{document}